\documentclass{article}
\usepackage[cp1251]{inputenc}
\usepackage[english]{babel}
\usepackage{graphicx}
\usepackage{amsmath}
\usepackage{amssymb}
\usepackage{latexsym}
\usepackage{natbib}
\newcommand*\apjs{ApJS }

\newcommand*\aap{A\&A }
\newcommand*\apjl{ApJL }

\begin{document}
\begin{flushright}
Journal-Ref: Astronomy Letters, 2020, Vol. 46, No. 11, pp. 774-782
\end{flushright}

\begin{center}
\Large {\bf Long-Term Dynamics of Planetesimals in Planetary Chaotic Zones}\\

\vspace{0.5cm}
\large {\bf T.V.\,Demidova$^1$, I.I.\,Shevchenko$^{2,3}$}

\normalsize
$^1$ Crimean Astrophysical Observatory, Russian Academy of Sciences, Nauchny, 298409 Russia\\
$^2$ St. Petersburg State University, Universitetskii pr. 28, St. Petersburg, 198504 Russia\\
$^3$ Pulkovo Astronomical Observatory, Russian Academy of Sciences,
Pulkovskoe sh. 65, St. Petersburg, 196140 Russia
\end{center}
\begin{center}
e-mail:proxima1@list.ru
\end{center}
\normalsize
\begin{abstract}
Extensive numerical experiments on the long-term dynamics of planetesimals near the orbits
of planets around single stars with debris disks have been carried out. The radial sizes of planetesimal clusters and the planetary chaotic zone as a function of mass parameter $\mu$ (planet--star mass ratio) have been determined numerically with a high accuracy separately for the outer and inner parts of the chaotic zone. The results obtained have been analyzed and interpreted in light of existing analytical theories (based on the planet--planetesimal mean motion resonance overlap criterion) and in comparison with previous numerical-experiment approaches to the problem. We show and explain how the stepwise dependence of
the chaotic zone sizes on $\mu$ is determined by the marginal resonances.

\end{abstract}

\section{Introduction}
The large-scale structures revealed in the images of protoplanetary disks are among the most intriguing discoveries of the outgoing decade. Such structures are most often concentric light and dark rings~\cite{2015ApJ...808L...3A, 2016ApJ...820L..40A,2016ApJ...829L..35T, 2016PhRvL.117y1101I, 2016ApJ...818L..16Z, 2016A&A...595A.112G, 2017A&A...600A..72F}. Such structures can be formed through the influence of a massive planet on the protoplanetary disk material~\cite[see, e.g.,][]{2012AJ....144...45K, 2015A&A...574A..68F, 2019ApJ...872..112V, 2016MNRAS.463L..22D, 2017ApJ...843..127D}. However, other mechanisms of their formation that do not require
the presence of a planet are also described in the literature; among them the following ones have been studied most comprehensively: the concentration of dust near the snow lines of the materials contained in the protoplanetary disk~\cite[see, e.g.,][]{2015ApJ...815L..15B, 2015ApJ...806L...7Z, 2016ApJ...821...82O, 2017ApJ...845...68P} the influence of a magnetic field, including the magnetorotational instability effects
in the disk~\cite[see, e.g.,][]{2009ApJ...697.1269J, 2014ApJ...796...31B, 2014ApJ...784...15S, 2015A&A...574A..68F, 2017MNRAS.468.3850S, 2017A&A...600A..75B} and the gravitational instability effects in the disk~\citep{2014ApJ...794...55T}. 

Similar ring-like structures were also detected during observations of debris disks around binary and single stars~\cite{1998ApJ...506L.133G, 2004ASPC..321..305A, 2005Natur.435.1067K, 2011ApJ...743L...6T, 2000ApJ...538..793K}. In this case, the planetesimal rings can be in resonances with a planet unresolvable in the disk images~\citep{2010ApJ...717.1123M, 2014AJ....148...59S}. The presence of the perturbing planet can limit the radial extent of the planetesimal disk both near the star and on the periphery~\citep{1999ApJ...527..918W, 2006MNRAS.373.1245Q, 2013ApJ...763..118S, 2014ApJ...780...65R}. 

On the other hand, populations of small bodies in $1:1$ resonance with the planets are well known in our Solar System; these are the so-called Trojan
asteroids. They exist at Jupiter~\citep[see, e.g.,][]{1999ssd..book.....M}, Mars~\cite{1990BAAS...22.1357B}, Neptune~\citep{2006Sci...313..511S}, Uranus~\cite{2013Sci...341..994A}, and the Earth~\citep{2011Natur.475..481C}. A dust ring coorbital with the Earth has been discovered~\cite{1994Natur.369..719D, 1995Natur.374..521R}.

\citet{1980AJ.....85.1122W} analytically established that the overlap of first-order particle--planet mean motion resonances (at which the orbital period of the planet is in ratio to the particle period as $p:(p+1)$ or $(p+1):p$, where $p=1, 2, 3, \dots$ ) is responsible for the formation of a ring-like chaotic zone in the radial neighborhood of the planet's orbit. Its radial sizes were subsequently estimated theoretically and numerically; if the mass parameter (planet--star mass ratio) $\mu << 1$, then the
radial half-width of the chaotic zone of a planet in a circular orbit is
\begin{equation}
\Delta a = C \mu^{\frac{2}{7}} a_\mathrm{p} ,
\end{equation}
where $a_p$ is the semimajor axis of the planet's orbit and the dimensionless numerical coefficient $C$ is determined by various authors to be within the range from $1.3$ to $1.6$~\citep{1989Icar...82..402D, 1998ASPC..149...37M, 2006MNRAS.373.1245Q, 2012MNRAS.419.3074M}.

The long-term evolution of planetesimals in the chaotic zone for the planets in circular orbits as a function of mass parameter was investigated by~\citet{2015ApJ...799...41M}; the sizes of the chaotic zone and the timescale of particle clearing due to their
escape from the system or infall onto the planet or the host star were estimated.

The properties of the chaotic zone for the planets
in elliptical orbits were considered by~\citet{2002ApJ...578L.149Q, 2003ApJ...588.1110K, 2006MNRAS.373.1245Q}. In the latter paper it was found that the sizes of the chaotic zone are virtually independent of the planet's eccentricity if its value is
less than $0.3$.

\section{The model and the simulation technique}
In this paper we simulated the dynamics of passively gravitating planetesimals in the gravitational field of a star--planet system. The physical sizes of the star and the planet are assumed to be zero. Thus,
the possibility of particle infall onto the planet or the star is ruled out; the chaotic zone clearing of particles occurs through their ejection from the system. The mass of the central star for all models was chosen
to be equal to the solar one $M_{\odot}$, while the mass of
the planet was varied in such a way that the planet--star mass ratio ($\mu$) changed from $\log\mu = -1.5$ to $-5.5$ with a $0.05$ step. The planet's orbit is circular with a period $P_\mathrm{p} = 1$ year. Initially, $10^4$ test particles were uniformly distributed (in radius and azimuth) in circular barycentric orbits in a 0.4--1.8~AU ring.Their dynamics was simulated on a time scale of $10^4$ years.

The equations of motion were integrated with the Bulirsch-Stoer integrator~\citep[see][]{1992nrca.book.....P}. The maximum admissible simulation error was assumed to be $\epsilon = 10^{-10}$; during the simulations the planet's semimajor axis was kept constant within the specified accuracy, the conservation of the Jacobi integral for all particles was also checked. When the Jacobi integral changed by more than $1\%$, $\epsilon$ dropped to its limit of $10^{-14}$. As it turned out, for some particles this limit is still insufficient for the Jacobi integral to be conserved. The non-conservation of the Jacobi constant is attributable to close particle--planet encounters. Violations arose at distances less than $10^{-7}a_\mathrm{p}$ (which is $\sim 15$ km if the semimajor axis of the planet's orbit is $a_\mathrm{p}\approx 1$~AU). The relative
number of such particles is small: in our simulations their number did not exceed $0.5\%$ of the total number of particles. They were excluded from the sample when analyzing the results.

The particles whose orbital semimajor axis during the simulations reached $2$~AU, were deemed to have escaped from the system and were excluded from
the simulations~\citep[similarly to the criterion adopted by][]{2015ApJ...799...41M}.

\section{The sizes of the chaotic zone}
\begin{figure}
\includegraphics[width=\columnwidth]{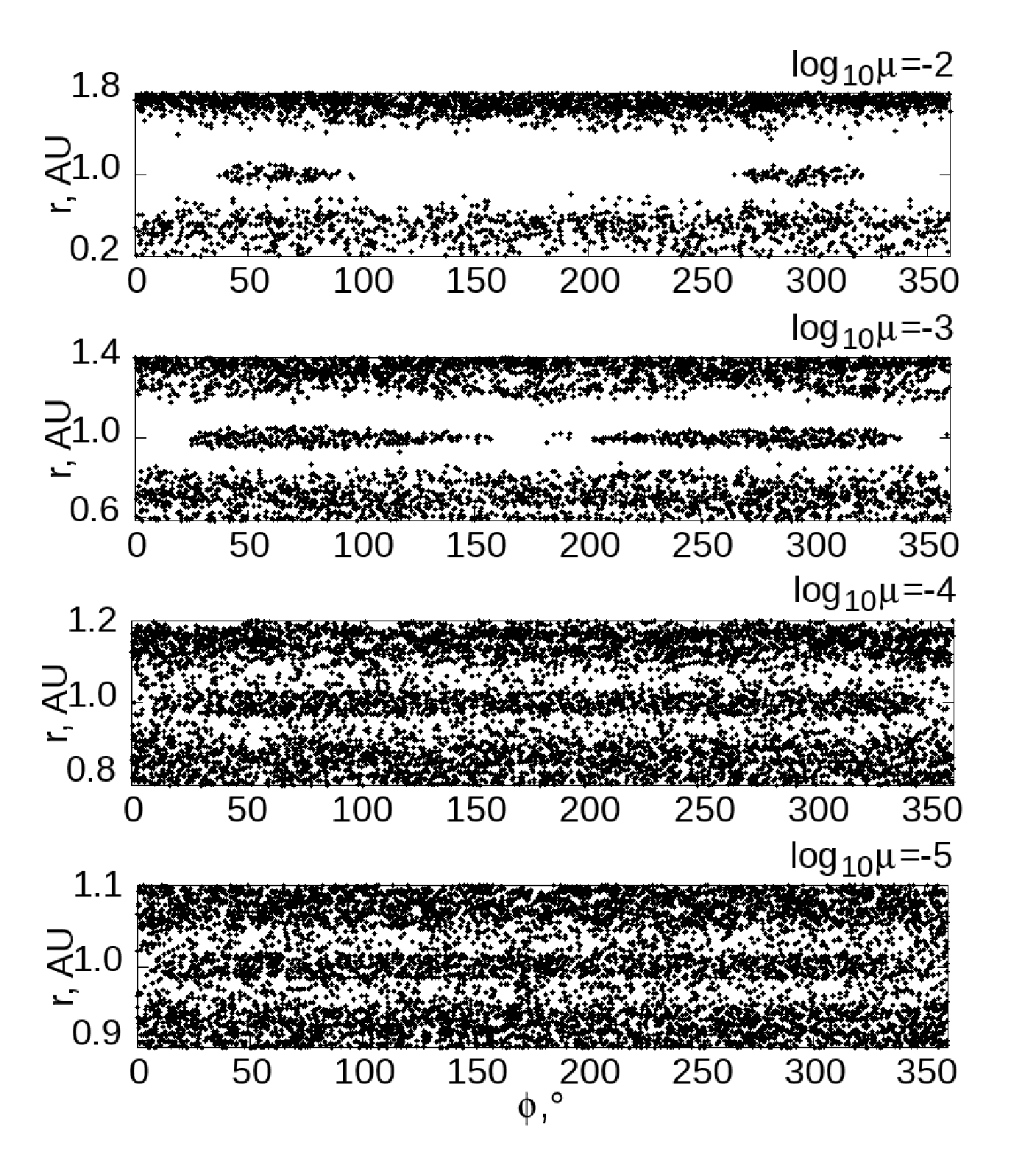}
\caption{The distribution of particles in the disk after $10^4$ revolutions of the planet in the polar barycentric system of coordinates $r$ (radial distance) and $\phi$ (azimuthal angle) revolving with the planet's orbital frequency; the planet is located at the point with $r=1$~AU, $\phi=0$.The logarithm of the planet--star mass ratio is specified above the graph.} \label{fig:part}
\end{figure}

Figure~\ref{fig:part} presents the final distributions of particles
near the planet's orbit in polar barycentric coordinates at a time of $10^4$ years from the beginning of evolution for four values of $\mu$. Obviously, the chaotic zone is cleared more poorly with decreasing mass parameter $\mu$; at the same time, the coorbital structures near the Lagrange points L4 and L5 bunch up and stretch along the planet's orbit, taking a horseshoe shape at $\log\mu < -3.1$. Our simulations showed that
the structure of the region is formed during the first  thousand orbital revolutions of the planet and, subsequently, its shape and sizes change only slightly.

\begin{figure}
\includegraphics[width=\columnwidth]{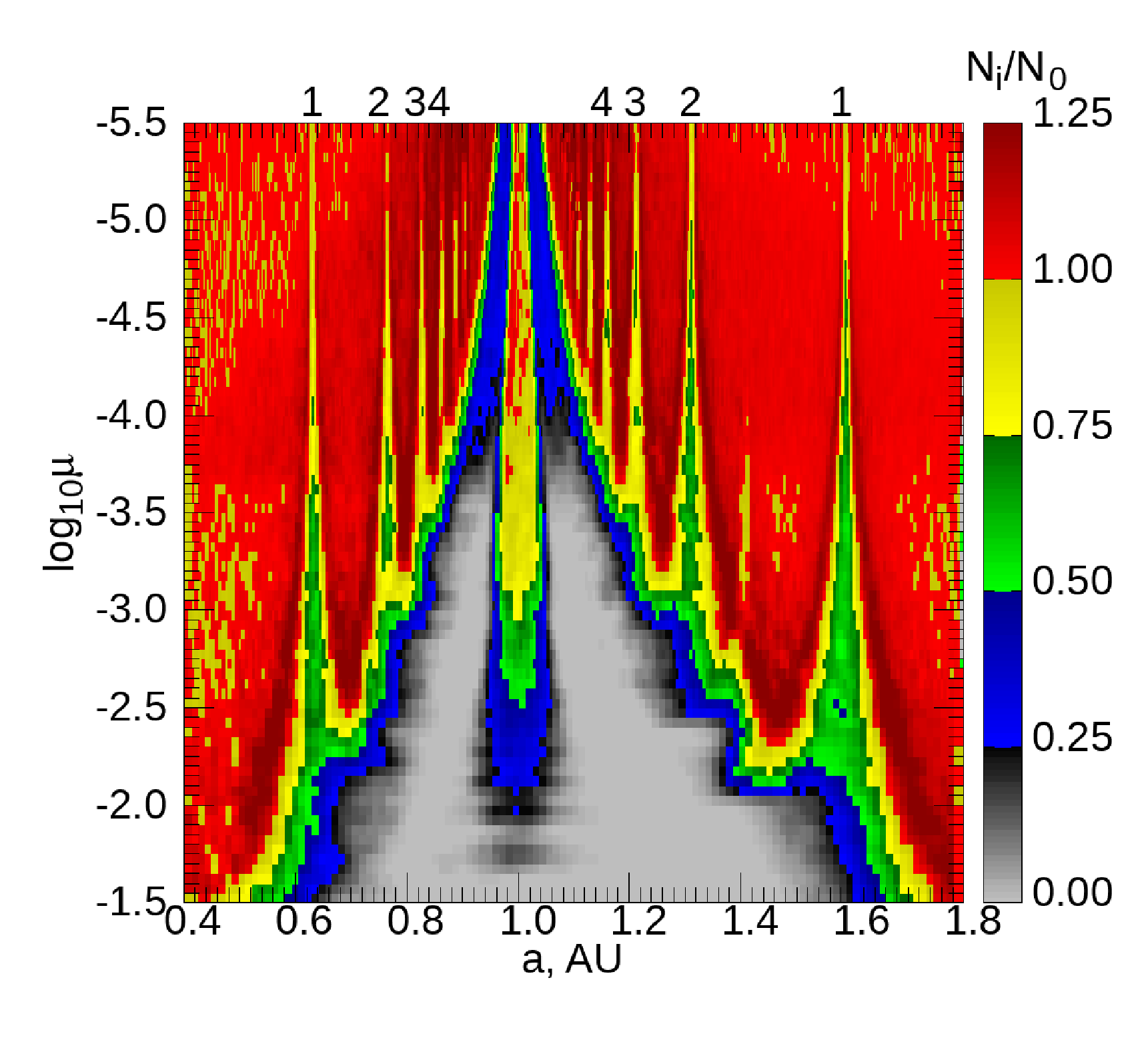}
\caption{The number of particles $N_i$ (with respect to the initial $N_0$) conserved in the disk after $10^4$ revolutions of the planet as a function of radial distance $a$ from the star (horizontal axis) and mass parameter $\mu$ (vertical axis). The numbers above the graphs specify the index $p$ for the $(p+1):p$ (at $a < 1$~AU) and $p:(p+1)$ (at $a > 1$~AU) resonances.}
\label{fig:ATm}
\end{figure}

To characterize the behavior of particles inside and near the planetary chaotic zone, the planetesimal disk was divided into $100$ The boundaries of the chaotic zone (both inner and outer ones) were determined by
the position of the ring with ${N_i}/{N_0}<50\%$. For each ring we computed the current number of particles ${N_i}$ contained in it as a function of time and its ratio to the initial one. Figure~\ref{fig:ATm} shows ${N_i}/{N_0}$ as a function of $\mu$ and distance from the system's center of mass at a time of $10^4$ years from the beginning of evolution;
here, $N_0$ is the initial number of particles in the ring. As follows from the graph, the change in the width of the cleared zone with decreasing $\mu$ is stepwise in pattern. At all $\mu$ the width of the chaotic zone
inside the planet's orbit is noticeably smaller than that outside. At $\mu \lesssim 0.01$ a structure coorbital with the planet begins to form; the relative amount of material in it increases with decreasing $\mu$.

\begin{figure}
\includegraphics[width=\columnwidth]{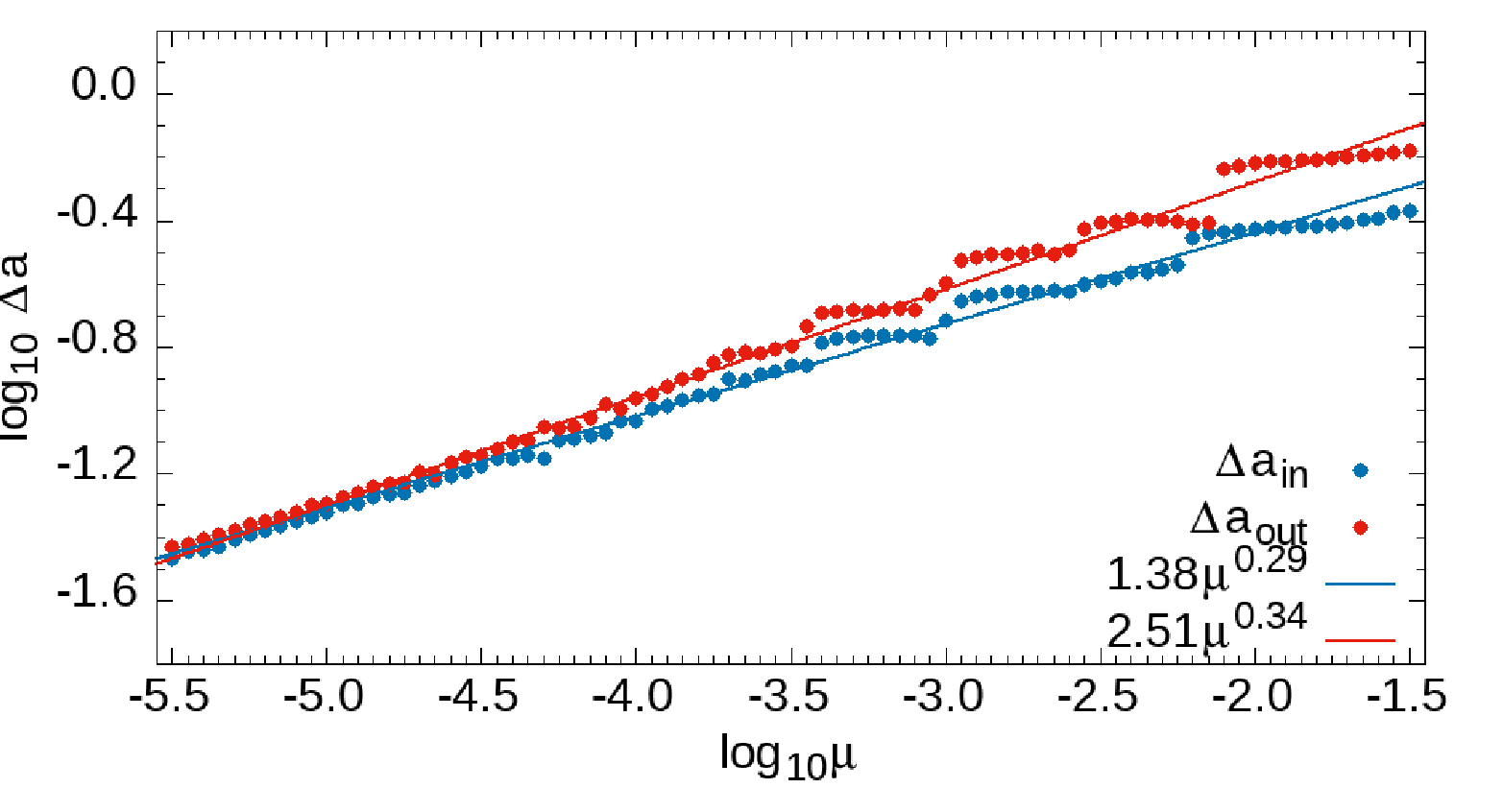}
\caption{Radial width $\Delta a$ of the inner (blue circles) and outer (red circles) parts of the chaotic zone versus mass parameter $\mu$, in a logarithmic scale. The dependences~(\ref{eq:ain}) and  (\ref{eq:aout}) are indicated by the blue and red lines, respectively; $\Delta a$ is expressed in units of the planet's semimajor axis $a_\mathrm{p}$.}
\label{fig:TAL0}
\end{figure}

In Fig.~\ref{fig:TAL0} the radial sizes of the inner ($\Delta a_\mathrm{in} = a_\mathrm{p} - a_\mathrm{in}$) and outer ($\Delta a_\mathrm{out} = a_\mathrm{out} - a_\mathrm{p}$) parts of the chaotic zone are shown as a function of $\mu$.

The dependence is stepwise due to the separation of resonances, for example, 2:1 (at $\log\mu = -2.2$ -- $-2.25$) and 1:2 (at $\log\mu=-2.1$ -- $-2.15$), 3:5 (at $\log\mu = -2.55$ -- $-2.6$), 3:2 and 2:3 (both at $\log
\mu = -2.95$ -- $-3$) , 4:3 and 3:4 (both at $\log\mu = -3.4$
-- $-3.5$) from the chaotic zone. The resonance separation is asynchronous in $\mu$ for its inner and outer parts. Although the curves for the
radial width are stepwise, they can be approximately described by the following simple relations:

\begin{equation}
\Delta a_\mathrm{in} = (1.38 \pm 0.04) \cdot \mu^{0.29\pm0.01}a_\mathrm{p}
\label{eq:ain}
\end{equation}

\noindent for the inner part of the chaotic zone (the blue line in Fig.~\ref{fig:TAL0}) and

\begin{equation}
\Delta a_\mathrm{out} = (2.51 \pm 0.08) \cdot \mu^{0.34 \pm 0.01}
a_\mathrm{p}
\label{eq:aout}
\end{equation}

\noindent for the outer part (the red line in Fig.~\ref{fig:TAL0})..

\begin{figure}
\includegraphics[width=\columnwidth]{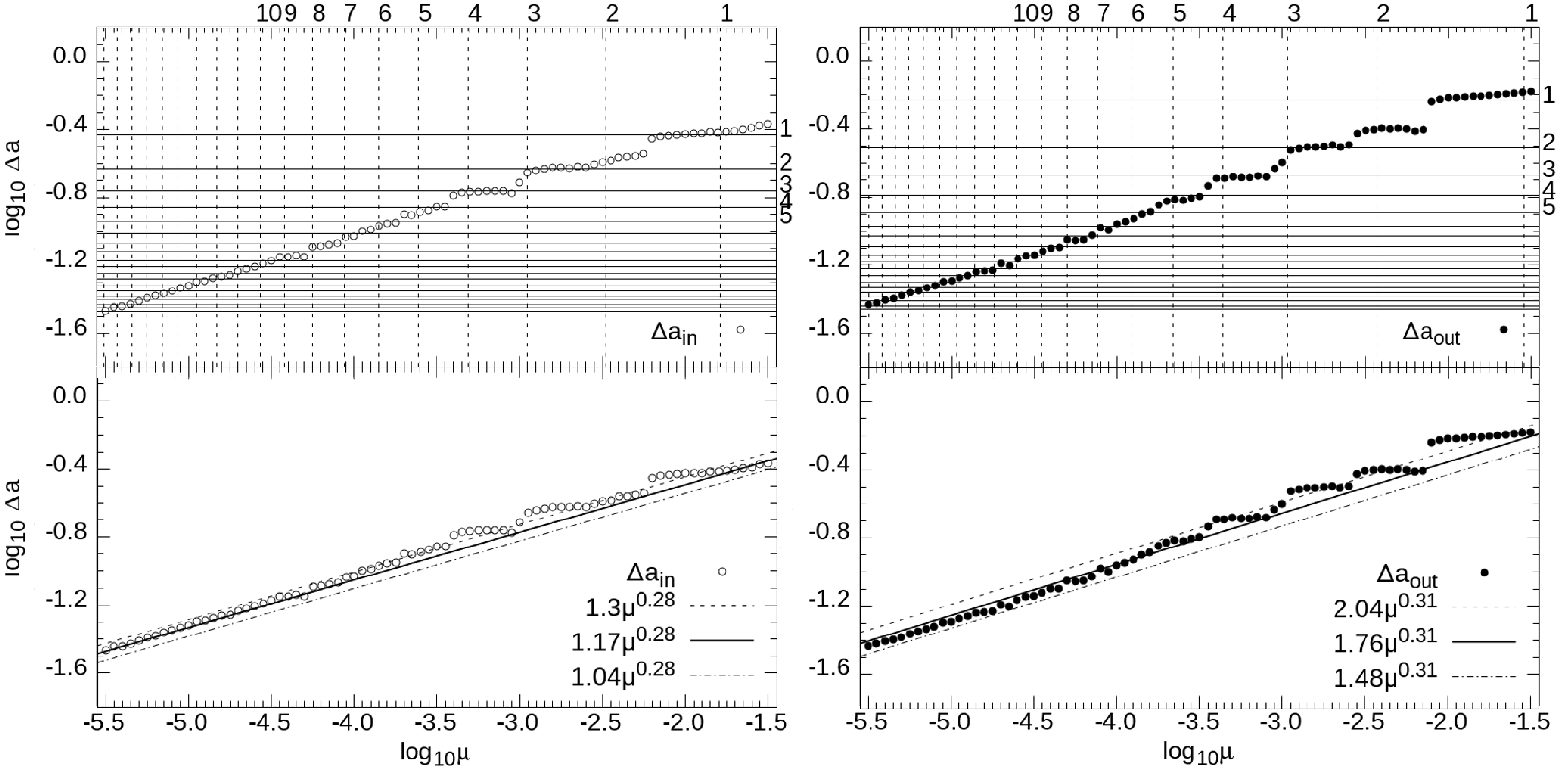}
\caption{The same as Fig.~\ref{fig:TAL0}, but on separate panels and with different fitting curves. On the upper panels the horizontal solid
lines indicate the positions of the first-order mean motion resonances; the vertical dashed lines indicate the values of $\log\mu$ at which the resonance position coincides with the average boundary of the chaotic zone. The resonance indices $p$ are specified on the panels at the top and on the right. On the lower panels the solid line indicates the average dependences on $\mu$ for the width of the inner and outer parts of the chaotic zone according to~\citet{2015ApJ...799...41M}, while the dotted and dash-dotted
lines indicate the limiting dependences including the error bar according to the same paper.} \label{fig:TAL4}
\end{figure}

The horizontal lines indicating the positions of the first-order resonances are additionally plotted on the upper panels of Fig.~\ref{fig:TAL4}, which presents the same simulation data as those in Fig.~\ref{fig:TAL0}: 

\begin{equation}
 \Delta a_\mathrm{in} =\Big[1 -
\Big(\frac{p}{p+1}\Big)^{\frac{2}{3}}\Big]a_\mathrm{p}
\end{equation}

\noindent (the inner boundary of the chaotic zone, the left panel) and
\begin{equation}
 \Delta a_\mathrm{out} = \Big[\Big(\frac{p+1}{p}\Big)^{\frac{2}{3}} - 1\Big]
a_\mathrm{p}
\end{equation}

\noindent (the outer boundary, the right panel). Here, $p = 1, 2, 3,
\dots 19$. It is obvious from the graphs that the abrupt changes in the width of the chaotic zone occur from resonance to resonance mostly between
the neighboring first-order resonances; however, between the 2 : 1 and 3 : 2 resonances (for which $p = 1$ and $p = 2$, respectively) the intermediate second order 5 : 3 resonance also clearly manifests itself. Thus, there is an intermediate step whose existence is attributable to the separation of the second-order resonance from the planetary chaotic zone; the latter
is also clearly seen in Fig.~\ref{fig:ATm}.

The vertical lines on the upper panels of Fig.~\ref{fig:TAL4} correspond to the values of $\mu$, at which the positions of the $p = 1, 2, 3, \dots $ resonances coincide with the boundary of the chaotic zone found from the best fits in Table~1 from~\citet{2015ApJ...799...41M}:

\begin{equation} \log\mu = \frac{1}{0.28} \log \Big( \frac{\Delta
a_\mathrm{in}}{1.17} \Big) \label{eq:MM_in}
\end{equation}

\noindent (the inner boundary, the left panel) and

\begin{equation} \log\mu = \frac{1}{0.31} \log \Big( \frac{\Delta
a_\mathrm{out}}{1.76} \Big) \label{eq:MM_out}
\end{equation}

\noindent (the outer boundary, the right panel). At $p > 1$ the vertical lines coincide with the jumps of the numerical curve. The non-coincidence at $p = 1$ is explained by the fact that near the marginal resonance with $p = 1$ the near-separatrix chaotic layer has a significant width and, therefore, its combination with the main layer is shifted in $\mu$ (Fig.~\ref{fig:ATm}).

The dependences (\ref{eq:MM_in}) and (\ref{eq:MM_out}) are additionally plotted on the lower panels of Fig.~\ref{fig:TAL4} with the same numerical
data. The upper and lower limiting dependences derived by taking into account the statistical errors of the coefficients in the formulas are also
shown (the dotted and dash--dotted lines). Our data lie within the limits of the error bars at $\log{\mu} < -3.5$, but they partially go beyond these limits at larger values of $\mu$. This is explained by the fact that in~\citet{2015ApJ...799...41M} the step of the models in $\log{\mu}$ was 0.5 and exceeded the one adopted by us by a factor of 10; therefore, the stepwise dependence could not manifest itself.

\begin{figure}
\includegraphics[width=\columnwidth]{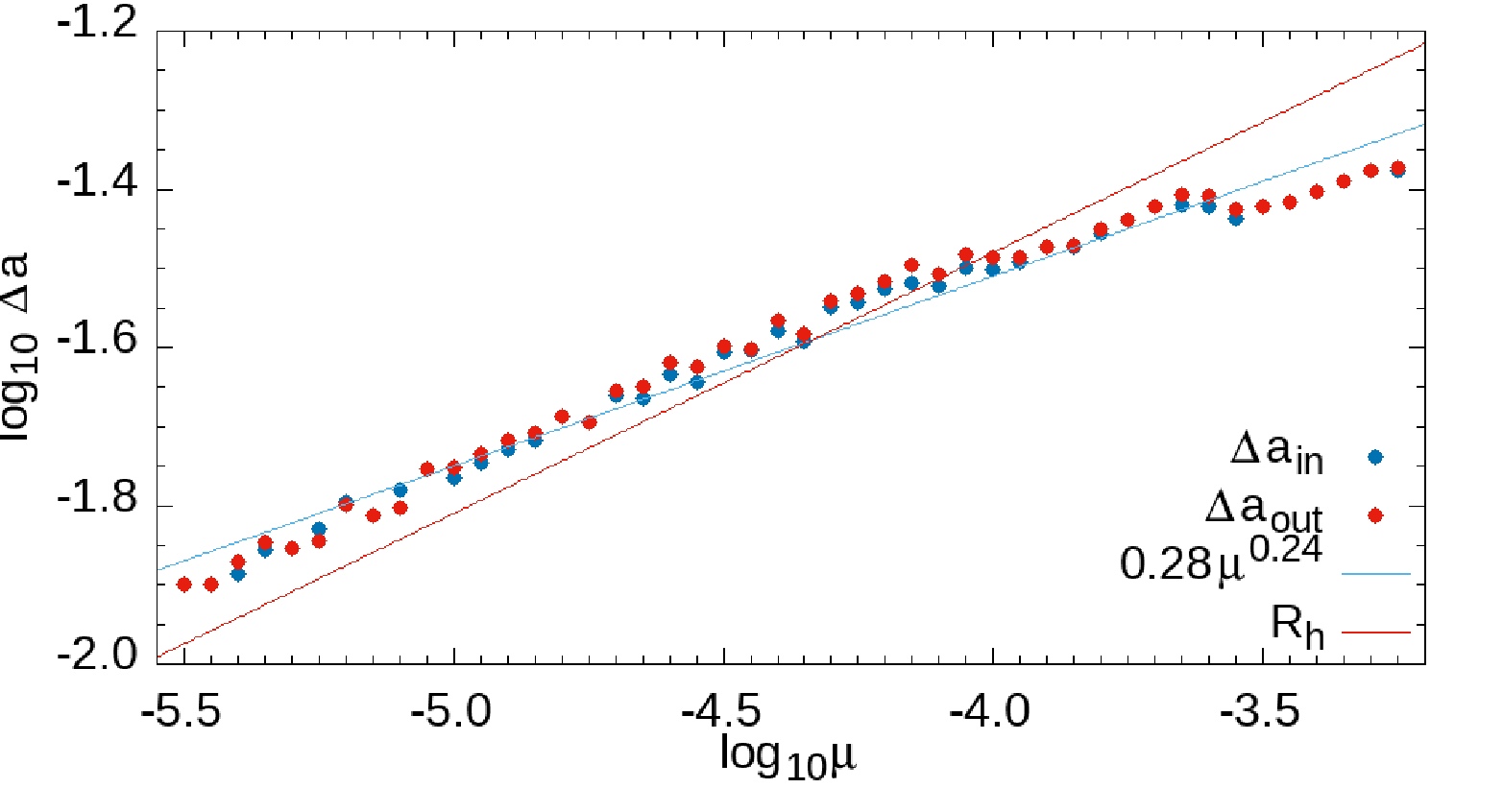}
\caption{Width of the inner (blue circles) and outer (red circles) parts of the coorbital swarm of particles versus mass parameter $\mu$ in a logarithmic scale. The dependence~(\ref{eq:ar}) and the dependence of the Hill radius ($R_h=(\mu/3)^{1/3}$) on $\mu$ are indicated by the
blue and red lines, respectively. $\Delta a$ is expressed in units of the planet's semimajor axis $a_\mathrm{p}$.}
\label{fig:RW}
\end{figure}

Similarly, we computed the radial boundaries of the coorbital swarm. It is prominent at $\log{\mu} < -3$ (Fig.~\ref{fig:ATm}). The sizes of the inner ($\Delta a^r_\mathrm{in} = a_\mathrm{p} - a^r_\mathrm{in}$) and outer ($\Delta a^r_\mathrm{out} = a^r_\mathrm{out} - a_\mathrm{p}$) parts of the coorbital swarm as a function of $\mu$ are shown in Fig.~\ref{fig:RW}. It follows from the graph that the dependences for the two parts of the swarm virtually coincide; thus, this horseshoe structure is radially symmetric relative to the planet's orbit. The general averaged dependence is

\begin{equation}
\Delta a^r = (0.28 \pm 0.03) \cdot \mu^{0.240 \pm 0.006} a_\mathrm{p}.
\label{eq:ar}
\end{equation}

\noindent This relation is similar in form to the one derived by~\citet{2018SoSyR..52..180D}; here, the coefficients were calculated
with a greater accuracy.

\section*{Marginal resonances}

The concentration of planetesimals drops, as a result of their escape, not only in the main (considered above) planetary chaotic zone, where the orbital resonances overlap, but also in the radial neighborhoods of the resonances located near it~\citep{2016MNRAS.463L..22D,2016ApJ...818..159T}, because the separatrices of close resonances can be strongly perturbed and, in addition, their subresonances can overlap inside the subresonance multiplets~\citep{2020ASSL..463.....S}. The main planetary chaotic zone and the resonance zones located near it can constitute more or less prominent three-lane and even multi-lane structures.

\begin{figure}
\includegraphics[width=\columnwidth]{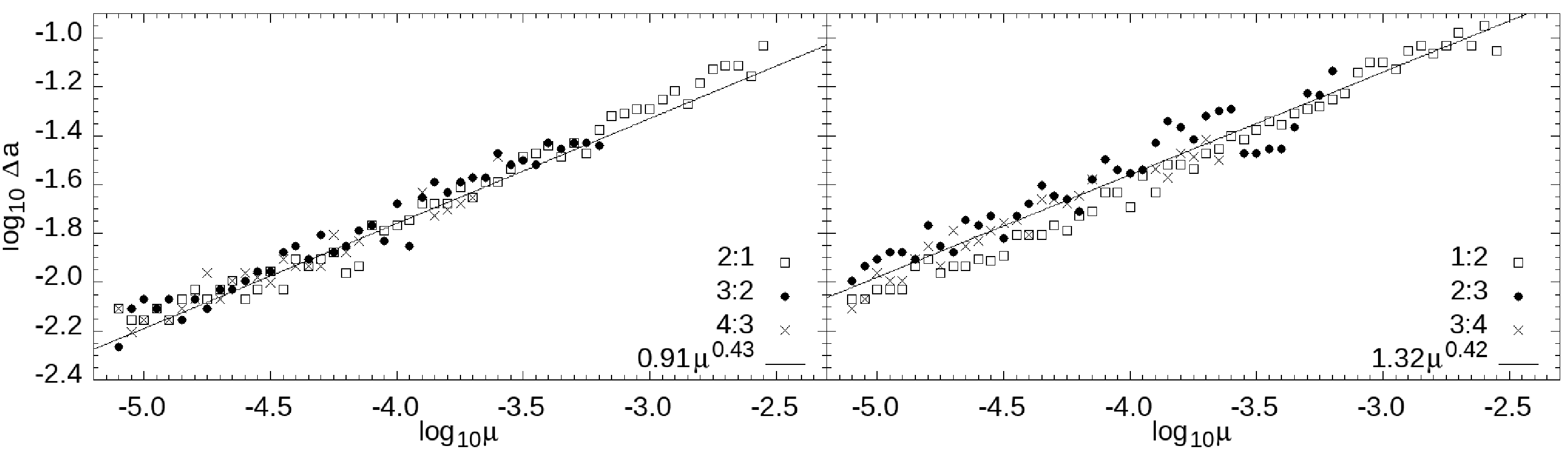}
\caption{The width of the region of reduced particle concentration near the resonances separating away from the chaotic zone. The right panel is for the inner (relative to the planet's orbit) resonances, the solid line indicates the dependence~(\ref{eq:api}); the left panel is for the outer resonances, the solid line indicates the dependence~(\ref{eq:apo}). $\Delta a$ is expressed in units of the planet's semimajor axis $a_\mathrm{p}$.}
\label{fig:pg}
\end{figure}

An insight into the pattern and radial width of the ``secondary'' chaotic zones attributable to the resonances that are located near the planetary chaotic zone, but do not merge with it, can be gained by specifying
the horizontal sections $\mu=const$ on the diagram presented in Fig.~\ref{fig:ATm}. The width of the secondary zones is maximal when the resonances responsible for them are separated from the main chaotic zone.
We numerically determined the width of the regions with a reduced particle concentration near the resonances moving away from the planetary chaotic zone as $\mu$ decreases. In Fig.~\ref{fig:ATm} the positions of the most
important resonances are marked by the numbers ($p = 1, 2, 3,4$) above the graph. The boundaries of the cleared resonance zones are determined by the condition ${N_i}/{N_0} < 1$. Our simulations showed that the width of the zones is virtually independent of $p$ (at $p = 1, 2, 3$) and depends on $\mu$ as follows: 

\begin{equation}
\Delta a^p_\mathrm{in} = 0.91^{+0.08}_{-0.06} \ \mu^{0.43 \pm 0.01} a_\mathrm{p}.
\label{eq:api}
\end{equation}

\noindent for the $(p+1)$:$p$ resonances (the inner ones relative to the planet's orbit, see Fig.~\ref{fig:pg}, the right panel) and

\begin{equation}
\Delta a^r_\mathrm{out} = 1.32^{+0.14}_{-0.13} \ \mu^{0.42 \pm 0.01} a_\mathrm{p}.
\label{eq:apo}
\end{equation}

\noindent for the $p$:$(p+1)$ resonances (the outer ones relative to the planet's orbit, see Fig.~\ref{fig:pg}, the left panel).

The sizes of the chaotic zone decrease sharply (in a stepwise manner) when the next marginal resonance is separated from the boundary of the chaotic layer as $\mu$ decreases. There are especially abrupt jumps when the first-order resonances are separated, which is obvious from Fig.~\ref{fig:TAL4}. This phenomenon is analogous to the separation of marginal resonances from the boundary of the chaotic zone of nonlinear resonance in the perturbed pendulum model~\citep[see][]{1998PhyS...57..185S, 2008PhLA..372..808S,2012PhRvE..85f6202S}. Thus, the stepwise dependence of the sizes of the chaotic planetary zone on $\mu$ is determined by the marginal resonances, primarily by the first order $(p+1)$:$p$ resonances, where $p=1, 2, 3, \dots$.

\section*{Discussion of results: observational aspects}

Ring-like structures are often present in the images of circumstellar debris disks~\citep{1998ApJ...506L.133G, 2009ApJ...693..734C,2011ApJ...743L...6T,2012AJ....144...45K}. However, observations of a single ring do not allow the orbit and mass of the ringcontrolling planet to be judged. Observationally, an analysis of multi-lane structures can be useful for determining the planet's parameters~\citep{2016MNRAS.463L..22D} by assuming that the resonances
are responsible for the formation of the lanes.

The relations derived above can be applied when analyzing the structures in which a bright (B) ring is in resonance with one or more dark (D) rings (B-
D, D-B, D-B-D structures) or when analyzing the structures in which two dark rings are in resonance (D-D structures) if the coorbital swarm is indistinct
and unresolvable in observations; see the nomenclature of lanes in~\citet{2016MNRAS.463L..22D}. Note that in the model with a circular planetary orbit adopted by us, according to the performed simulations,
the coorbital swarm of particles takes a horseshoe (almost closed ring-like) shape at $\log\mu < -3.1$. 

It should be emphasized that here we disregard the gravitational mutual scattering of planetesimals, their collisions and interactions with the gas. It was shown in~\citet{2018AstL...44..119D} that the pairwise relaxation timescale of planetesimals with a  size $\sim 10$~km and a density of $2$~g~cm$^{-3}$ in a typical disk is limited from below by $10^7$ years. Based on Eq. (1) from~\citet{2019AJ....157..202S}, we can estimate the collision timescale of such planetesimals: at a distance of $1$~AU from the star it is $\sim 5 \times 10^4$ years. By assuming that the mass of the disk gas is about 1\% of the solar mass and that the material is distributed according to the laws described, for example, in~\citet{2019ApJ...887L..15D}, we can estimate the time it takes for the planetesimal velocity to decrease by a factor of $e\approx 2.7$~\citep{1977MNRAS.180...57W}. At a gas density of $1.6 \times 10^{-10}$~g~cm$^{-3}$ and a sound speed of $0.8$ km s$^{-1}$ for such planetesimals this time is $\sim 3 \times 10^4$ years. Thus, on the timescales adopted in our simulations these effects can most likely manifest themselves only weakly, but they can be significant on longer time scales.

\citet{2018ApJ...869L..42H} and \citet{2019ApJ...872..112V} established that some of the ring structures observed in the ALMA images of gas-dust disks reside in orbital resonances relative to one another. These images characterize the thermal radiation from small dust particles of millimeter and centimeter sizes; the dynamics of such particles can be strongly affected by the disk gas. The influence of resonances with the
planet on the formation of structures in such disks can be significant; however, in future, simulations including the interaction of the disk gas and dust components as well as the production and redistribution
of small particles in the gas are needed to describe the structures.

\section*{Conclusions}

We carried out extensive numerical experiments on the long-term dynamics of planetesimals near the orbits of planets around single stars with debris disks. The radial sizes of planetesimal clusters and a planetary
chaotic zone as a function of mass parameter $\mu$ (planet--star mass ratio) were numerically determined with a high accuracy. Our simulations were
performed separately for the outer and inner parts of the chaotic zone.

The results obtained were analyzed and interpreted in light of existing analytical theories and in comparison with previous numerical-experiment
approaches to the problem. The width of the chaotic zone was found to change abruptly with decreasing $\mu$ due to the separation of the particle--planet mean motion resonances from the zone. The stepwise dependence of the chaotic zone sizes on $\mu$ is determined by the marginal resonances.

We also showed that the concentration of planetesimals could be reduced significantly near the resonances outside the planetary chaotic zone not only in the case of ``main'' resonances with $p=1$~\citep[as established
previously by ][]{2016MNRAS.463L..22D,2016ApJ...818..159T}, but also
for resonances with $p>1$. For resonances with $p=1$ we constructed the dependences on mass parameter  $\mu$ for the width of the cleared resonance zones. The width of the zones was found to be greater in the case of outer (relative to the planet's orbit) resonances.

\textbf{Acknowledgments}. We are grateful to the referees for useful remarks.

\textbf{Funding}.
This work was supported by grant no. 075-15-2020-780 ``Theoretical and Experimental Studies of the Formation and Evolution of Extrasolar Planetary
Systems and the Characteristics of Exoplanets'' of the Ministry of Science and Higher Education of the Russian Federation.

\bibliography{biblio}
\end{document}